\documentclass[aps,prr,floats,floatfix,twocolumn,fleqn]{revtex4-1}

\usepackage{graphicx}
\usepackage{epstopdf}
\graphicspath{{./Figs/}}

\usepackage[percent]{overpic} 

\usepackage{latexsym}
\usepackage{amsmath}
\usepackage{amssymb}
\usepackage{bm}
\usepackage{wasysym}

\usepackage{ifthen}
\usepackage{color}
\usepackage[colorlinks,allcolors=blue,bookmarks=true]{hyperref}

\newcommand{\mass}{\mathsf{m}}

\newcommand{\amatrix}[1]{\begin{matrix} #1 \end{matrix}}

\newcommand{\bra}[1]{\left\langle #1 \right|}
\newcommand{\ket}[1]{\left| #1 \right\rangle}
\newcommand{\braket}[1]{\left\langle #1 \right\rangle }
\newcommand{\avg}[1]{\left\langle #1 \right\rangle }
\newcommand{\kb}[2]{ | #1 \rangle \langle #2 |}

\newcommand{\Braket}[2]{\left\langle #1 \middle| #2 \right\rangle}
\newcommand{\BraKet}[3]{\left\langle #1 \middle| #2 \middle| #3 \right\rangle}

\newcommand{\beq}{\begin{eqnarray}}
\newcommand{\eeq}{\end{eqnarray}}

\newcommand{\hide}[1]{}  

\newcommand{\Eq}[1]{{\textcolor{blue}{Eq.}}~\!\!(\ref{#1})} 
\newcommand{\Sec}[1]{{\textcolor{blue}{Sec.}}~(\ref{#1})} 
\newcommand{\App}[1]{{\textcolor{blue}{Appendix}}~(\ref{#1})} 
\newcommand{\Fig}[1] {{\textcolor{blue}{Fig.}}~\!\!\ref{#1}}
\newcommand{\sect}[1]{{\bf #1.-- }}

\newcommand{\hrefl}[1]{\href{#1}{[link]}}

\begin{document}

\title{Breakdown of quantum-to-classical correspondence \\ for diffusion in high temperature thermal environment}

\author{Dekel Shapira, Doron Cohen}

\affiliation{
\mbox{Department of Physics, Ben-Gurion University of the Negev, Beer-Sheva 84105, Israel} 
}

\begin{abstract}
We re-consider the old problem of Brownian motion in homogeneous high-temperature thermal environment. The semiclassical theory implies that the diffusion coefficient does not depend on whether the thermal fluctuations are correlated in space or disordered. We show that the corresponding quantum analysis exhibits a remarkable breakdown of quantum-to-classical correspondence. Explicit results are found for a tight binding model, within the framework of an Ohmic master equation, where we distinguish between on-site and on-bond dissipators. The breakdown is second-order in the inverse temperature, and therefore, on the quantitative side, involves an inherent ambiguity that is related to the Ohmic approximation scheme.      
\end{abstract}

\maketitle


\section{Introduction}

The fingerprints of quantum mechanics on Brownian motion is an intriguing theme \cite{Caldeira1983,CALDEIRA1983374,HakimAmbegaokar1985}. 
This theme concerns also the motion of a particle or an exciton on a lattice
\cite{MadhukarPost1977,Weiss1985,Kumar1985,aslangul1986quantum,Weiss1991,Dibyendu2008,Amir2009,
lloyd2011quantum,Moix_2013,CaoSilbeyWu2013,Kaplan2017ExitSite,Kaplan2017B,
dekorsy2000coupled,dubin2006macroscopic,nelson2018coherent}, 
or the closely related studies of motion in a washboard potential \cite{Schmid1983,Fisher1985QuantumBrownianPeriodic,Fisher1985QuantumBrownianPeriodic,AslangulPeriodicPotential1987}.

The traditional paradigm is/was that the effects of quantum mechanics show up only at low-temperatures, where non-classical effects are related to the failure of the Markovian approximation. This view has been challenged by publications regarding excitation transport in photosynthetic light-harvesting complexes, most notably by the experiment in \cite{engel2007evidence}, and by many theoretical publications
\cite{amerongen2000photosynthetic,ritz2002quantum,FlemingCheng2009,plenio2008dephasing,Rebentrost_2009,Alan2009,Sarovar_2013,higgins2014superabsorption,celardo2012superradiance,park2016enhanced}.
But by now it has been argued \cite{tiwari2013electronic,Tempelaar2014,Duan2017,Maiuri2018,Thyrhaug2018,QuanBioRevCao2020} that the transport there, by itself, is ``classical" in nature. 

Nevertheless, contrary to the traditional paradigm, 
we suggest below that quantum manifestation in stochastic motion can be detected via the high-temperature dependence of the transport coefficients. This opens a new avenue for challenging the traditional (classical) paradigm of Brownian motion.

\subsection{Brownian motion}

High temperature  ($T$) classical Brownian motion is described by 
the Langevin equation 
%
\beq \label{eq:langevin-p}
\dot{p} \ = \ -\eta \dot{x} + f,
\eeq
where $f$ is white noise of intensity $\nu$,  
related to the friction coefficient 
via the fluctuation dissipation relation  
\beq \label{eFDR}
\nu = 2 \eta T,  \ \ \ \ \
\text{[can be used as definition of $T$]}
\eeq
For the standard dispersion relation ${\dot{x}=(1/\mass)p}$, 
where $\mass$ is the mass of the particle, 
one obtains the following simple results
for the transport coefficients:
\beq \label{eMu}
\mu &=& \frac{1}{\eta}, \ \ \ \ \ \text{[mobility]} 
\\ \label{eDcoef}
D &=& \frac{T}{\eta}, \ \ \ \ \ \text{[diffusion coefficient]}
\eeq
The mobility $\mu$ is used to determine the drift velocity 
due to an applied bias; while $D$ is the coefficient 
that enters Fick's law.
The Einstein relation ${D/\mu =T}$ is satisfied.
It is important to realize that \Eq{eFDR} 
characterizes the thermal environment, 
while the Einstein relation characterizes the 
dissipative dynamics of the particle.

\subsection{Quantum signature}

One wonders whether the dependence of the transport 
coefficients ($\mu$ and $D$) 
on the dissipation parameters ($\eta$ and $\nu$) is universal. 
This is the main question that motivates the present study.

\sect{Common wisdom}
The high temperature noise arises from a fluctuating potential, 
namely, 
\beq
f \ \ =  \ \ -\partial_x \mathcal{U}(x,t)
\eeq
This potential features in general a spatial correlation scale~$\ell$.
Semiclassicaly, the transport coefficients do not depend on $\ell$, and the common practice,
as in the Caldeira-Leggett model \cite{Caldeira1983,CALDEIRA1983374}, 
is to assume that $f$ is independent of~$x$, meaning that  $\ell{=}\infty$.
But in the quantum treatment $\ell$ does show up in the analysis, 
because it determines the lineshape of the stochastic kernel $\mathcal{W}(k|k')$ 
for scattering from momentum~$k'$ to momentum~$k$.  
Namely, the width of the kernel (${\sim} 2\pi/\ell$)  
has implication on the transient decoherence process \cite{Cohen97Brownian,Cohen1997,EspositoGaspard2005}.  
Yet, one does not expect that this lineshape will have any effect 
on the long time spreading. The argument is simple: 
on the basis of the  central limit theorem successive convolutions 
should lead to a result that does not depend on the $\ell$-dependent 
lineshape of the stochastic kernel, but only on its second moment, 
which is characterized by~$\nu$.  
Consequently robust quantum-to-classical correspondence (QCC) is expected at high temperatures.
Such QCC can be regarded as an implication of the Thomas-Reiche-Kuhn-Bethe-Wang sum rule \cite{WangSumRule99}, 
or as an extension of the {\em restricted} QCC principle \cite{Cohen99DissipationChaotic,StotlandCohen2006}.   
%

\sect{Main Statement}
In the present work we show that~${\ell}$~independence of the transport 
coefficients ($\mu$ and $D$) is a fallacy. Given $\eta$ we shall see that~$D$ 
acquires a non-universal dependence on the temperature, 
that constitutes a quantum-mechanical signature.

\subsection{Tight binding model}  
\label{sec:tb}

Here we consider a particle or a single exciton that can hop along 
a one-dimensional chain whose sites are labeled by an integer index~$x$.
The dynamics of the isolated system is determined by the Hamiltonian 
\beq
\label{eq:H-tb-1}
\bm{H}^{(c)} \ \ = \ \ -c \cos(a\bm{p}) - f_0 \bm{x}  
\eeq
where $a$ is the lattice constant, and $c$ is the hopping frequency,
and $f_0$ is an applied bias.
The operators $e^{\mp i a\bm{p}}$ generate one-site displacements.
This Hamiltonian is of quantum mechanical origin, 
but may be treated semiclassically, by deriving 
the equations of motion ${\dot{\bm{x}} = ca\sin(a \bm{p})}$ and  ${\dot{\bm{p}}=f_0}$.
Adopting  the standard jargon of Condensed Matter textbooks, 
we shall call this {\em semiclassical} treatment of the dynamics.
The exact {\em quantum} dynamics of \Eq{eq:H-tb-1} is obtained 
from the Schrodinger equation \cite{Hartmann_korsch_2004} 
or equivaelently from the Lionville-von-Neumann equation for the 
probability matrix~$\rho$.

The dynamics on the lattice features the dispersion relation  ${\dot{\bm{x}} = v(\bm{p})}$, 
where ${v(p) = ca\sin(a p)}$. The continuum limit (small~$ap$)  
leads to the standard dispersion relation ${v=(1/\mass)p}$ 
with ${ \mass = 1/(c a^2) }$. Therefore we can regard the latter case 
as a  special regime of the former.   
Irrespective of the dispersion relation, if the particle is coupled 
to a thermal environment, the semiclassical treatment 
leads to ${\dot{\bm{p}}=F(t)}$, where the force contains 
a stochastic {\em noise} term and a {\em friction} term, 
namely, ${ F(t) =  f_0 + f(t) -\eta \dot{x} }$. 
In the absence of external bias ($f_0{=}0$) 
this leads to the Langevin equation \Eq{eq:langevin-p}.

In the corresponding high-temperature Markovian {\em quantum} treatment 
the dynamics is given by a master equation for the probability matrix \cite{Breuer2002}. 
This master equation incorporates extra term, aka dissipator,
that represent the noise and the friction: 
\beq \label{e1}
\frac{d\rho}{dt} \ = \ \mathcal{L} \rho \ = \ 
-i[\bm{H}^{(c)},\rho] + \mathcal{L}^{(\text{bath})} \rho
\eeq
The dissipator $\mathcal{L}^{(\text{bath})}$ is determined by the coupling 
between the isolated chain and the environment, 
and depends on the temperature of the bath.

\subsection{Regime diagram}

\begin{figure}

\begin{overpic}
{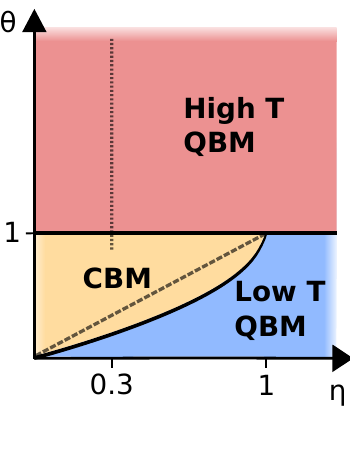}
\put(03,105){(a)}
\end{overpic}
\begin{overpic}
{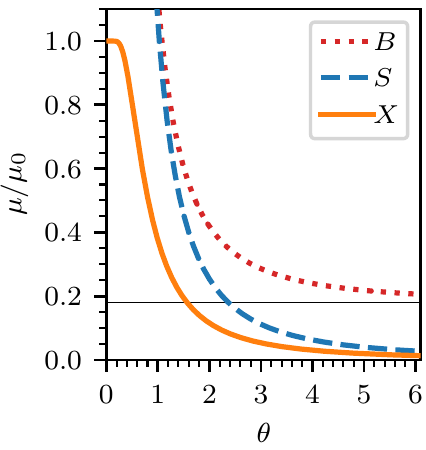}
\put(20,105){(b)}
\end{overpic}

\caption{\label{fg2}
{\bf The Brownian Motion regime diagram.} 
(a) The various regions in the  $(\eta,\theta)$ diagram are indicated. 
We distinguish between the Classical-like Brownian Motion (CBM) region; 
the low-temperature QBM region where memory effects dominates;
and the high-temperature QBM region that is discussed in this work. 
Note that below the dashed diagonal line (${\beta>1}$) memory effects should be 
taken into account.        
(b) The scaled mobility $\mu/\mu_0$ where ${\mu_0=1/\eta}$ versus~$\theta$, 
based on the analytical results that have been obtained for diffusion in the X/S coupling schemes. 
The result is independent of $\eta$. We also add the result for the B coupling scheme 
that approaches the finite asymptotic value ${\mu_{\infty}=2\eta}$ (horizontal line). 
In the latter case ${\eta=0.3}$ has been assumed. 
Note that the S/B results are applicable only in the ${\theta>1}$ regime. 
}
\end{figure}

Disregarding the optional applied bias~$f_0$,
the isolated tight binding model has no free parameters 
(formally we can set the units of time and length such that ${c=a=1}$).  
With bath, the continuum-version of Quantum Brownian Motion (QBM) 
features a single dimensionless parameter, 
the scaled inverse temperature $\beta$, which is the ratio between 
the thermal time $1/T$ and damping time ${\mass/\eta}$.
In the lattice problem one can define two dimensionless parameters
%
\beq
\alpha = \frac{\eta a^2}{2\pi}, 
\hspace{2cm} 
\theta=\frac{T}{c}
\eeq
Accordingly $\beta = \alpha/\theta$. 
In our model we set the units such that ${a=1}$, 
hence, disregarding $2\pi$ factor, 
our scaled friction parameter~$\eta$ is the same as~$\alpha$.
The regime diagram of the problem is displayed in \Fig{fg2}, 
and further discussed below. It contains both Classical-like Brownian Motion (CBM) regime, 
where memory effects are either not expressed or appear as a transient, 
and QBM regimes where the dynamics is drastically different.

\subsection{Relation to past studies}

The standard analysis of QBM \cite{HakimAmbegaokar1985} reveals 
that quantum-implied memory effects 
are expressed in the regime ${\beta > 1 }$, 
where a transient $\log(t)$ spreading is observed
in the absence of bias, followed by diffusion.

The later quantum dissipation literature, 
regarding the two-site spin-boson model \cite{LeggettEtAlDynamicsTwoLevel1987} 
and regarding multi-site chains \cite{aslangul1986quantum,AslangulPeriodicPotential1987},  
is focused in this low temperature regime 
where a transition from CBM-like behavior 
to over-damped or localized behavior is observed,
notably for large~$\alpha$ of order unity.   

Our interest is focused in the ${\alpha, \beta \ll 1 }$ regime.
This regime is roughly divided into two regions by the line ${\theta \sim 1}$, see \Fig{fg2}. 
Along this line the thermal de-Broglie wavelength 
of the particle is of order of the lattice constant, 
hence it bears formal analogy to the analysis
of QBM in cosine potential \cite{Fisher1985QuantumBrownianPeriodic}, 
where it marks the border to the regime where activation mechanism comes into action. 
In our tight binding model we have a single band, 
hence transport via thermal activation is not possible. 
Rather, in the ${\theta > 1}$ regime the momentum distribution within the band is roughly flat.
To avoid miss-understanding, what we call in the present study ``high temperature" regime 
assumes a single band approximation by construction.

\subsection{Outline}

Overview of the main results is presented in \Sec{sec:overview}.
The Ohmic master equation is explained in \Sec{sec:ohmic-dis}.
The semiclassical analysis is detailed in  \Sec{sec:semi-X} to \Sec{sec:semi-B}.
The quantum analysis is detailed in \Sec{sec:quantum-anal}.
The effective stochastic description is presented in \Sec{sec:stochastic},
where we discuss detailed-balance consideration as well.
Concise summary is provided in \Sec{sec:discussion}.

\section{Overview of main results}
\label{sec:overview}

In order to demonstrate that the temperature dependency of the
transport coefficients is $\ell$ dependent,
we consider in detail two extreme cases: 
{\bf (a)} The Caldeira-Leggett X{-}dissipator $\mathcal{L}^{(\text{X})}$
where a single bath is coupled to $\bm{x}$. 
This corresponds to non-disordered ($\ell{=}\infty$) bath.
{\bf (b)} The S{-}dissipator $\mathcal{L}^{(\text{S})}$ 
where each site is coupled to an independent bath. 
For this coupling $\ell{=}a$. 
In \Sec{sec:stochastic} we also present results for intermediate 
values of~$\ell$. 
For completeness we also consider another case:
{\bf (c)} The B{-}dissipator $\mathcal{L}^{(\text{B})}$ 
where each bond is coupled to an independent bath.
For all cases the dynamics is governed by the Ohmic  master equation \Eq{e1} 
and the dissipator $\mathcal{L}^{(\text{bath})}$
takes different forms according to the couplings.
For the 3 cases above the bath parameters are $\nu_i$ and $\eta_i$ with ${i=X,S,B}$.

\subsection{X-dissipation}

As a reference case we calculate the transport coefficients 
for a particle in a tight binding model, 
that is coupled to an Ohmic Caldeira-Leggett bath via the $x$ variable.
We term this standard case "X-dissipation". 
We set the length units such ${a=1}$.
The bath parameters $\nu_X$ and $\eta_X$ are chosen such that 
in the semiclassical \Eq{eq:langevin-p}, 
we have $\nu = \nu_X$ and $\eta = \eta_X$.
The result that we get for the diffusion coefficient is  
\begin{align}\label{eq:D-cl}
D^{\text{(X)}} &=  \left[1 - \frac{1}{[\mathrm{I}_0 (c/T)]^{2}}  \right] \dfrac{T}{\eta}
\end{align}
where $\textrm{I}_n$ is the modified Bessel function.
This result is exact to the extent that the (Markovian) Ohmic master equation can be trusted. 
For the mobility we get ${\mu = D/T}$ as expected for the Einstein relation.
A plot of the mobility versus temperature is provided in \Fig{fg2}.

\begin{figure}
\centering
\includegraphics[width=0.5\textwidth]{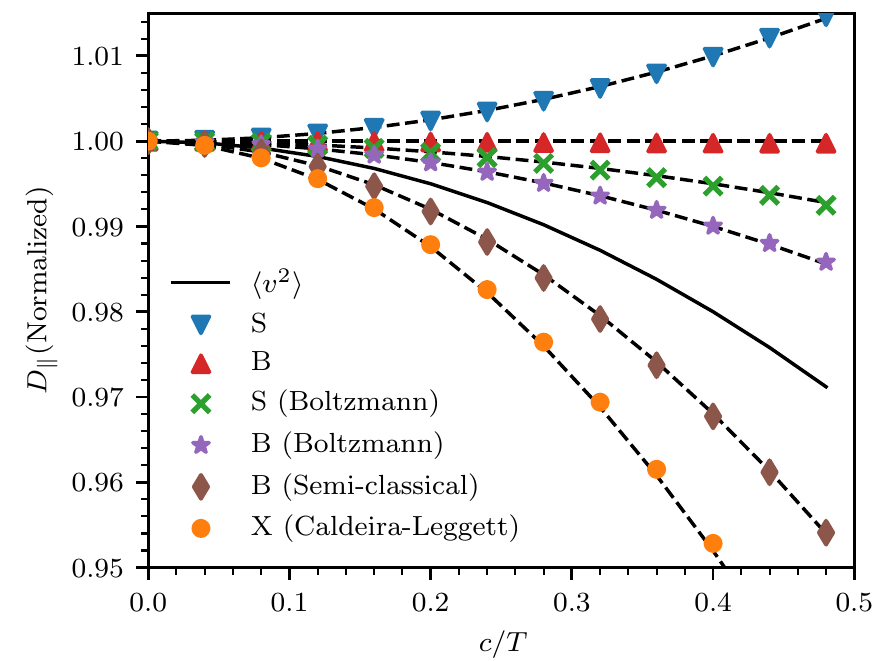}
\caption{\label{fg1}
{\bf Dependence of the diffusion coefficient on the temperature.}
Given~$\nu$ the coefficient $D_{\parallel}$ is plotted versus $(c/T)$ 
for the Ohmic master equation and different coupling schemes: 
Caldeira-Leggett~(X); Sites~(S); and Bonds~(B).  
The symbols are obtained numerically through an effective rate equation (see text). 
We also show results for the canonical (``Boltzmann") versions of the S/B master equation, 
and for the semi-classical result in the case of B-coupling  
(The semiclassical result for S-coupling is the same as X-coupling).  
The naive expectation~${D \propto  \braket{v^2}}$ is displayed for sake of comparison. 
}
\end{figure}

For low temperatures (in the sense ${T \ll c}$) one recovers the standard 
results \Eq{eMu} and \Eq{eDcoef} that apply for non-relativistic (linear) dispersion.
For high temperatures the result takes the form ${D^{\text{(X)}} = D_{\parallel}}$ with 
\beq \label{eDXS}
D_{\parallel} \ \ \approx \ C_{\parallel} \left[ 1 + A_{\parallel}  \left( \dfrac{c}{T} \right)^2 \right] \dfrac{c^{2}}{\nu}  
\eeq
where $C_{\parallel}{=}1$ and $A_{\parallel}{=}-5/16$.
The reason for using the subscript notation is clarified below.
The same expression appears for the S/B dissipators, 
with $\nu$ replaced by $\nu_S$ and $\nu_B$ respectively.
The dependence of $D$ on the temperature is plotted in \Fig{fg1}.
For sake of comparison we plot also the naive expectation ${D \propto  \braket{v^2}}$, 
with ${ v = c \sin(p) }$, where the average is over the canonical distribution.
This naive expectation would be valid, if the correlation time 
were independent of temperature (which is not the case).   
The high-temperature dependence is  
\beq \label{eq:vsqr}
\braket{v^2} \approx \left[ 1 + A  \left( \dfrac{c}{T} \right)^2 \right]\frac{c^2}{2}, 
\ \ \ \mbox{with $A=-1/8$} \ \ 
\eeq
In \Sec{sec:quantum-anal} we obtain the {\em same} result also within the framework 
of an {\em exact} quantum treatment.

\subsection{S-dissipation}

We shall explain that if the fluctuations of the bath are characterized 
by a finite correlation scale $\ell$, the semiclassical result for the transport 
coefficient are the same as for ${\ell=\infty}$. 
This rather trivial observation holds for any dispersion relation. 
Specifically for the tight binding model \Eq{eq:D-cl} is $\ell$ independent.
But this is not so in the quantum analysis. 
Here we analyze the other extreme limit of ${\ell \sim a}$, 
meaning that the bath fluctuations at different sites are uncorrelated.  
We obtain \Eq{eDXS} with  $A_{\parallel}{=}1/16$ and $\nu \mapsto \nu_S$,
which is not even the same sign when compared to the X-dissipation result.

\sect{Technical note:} 
For a tight binding model the parameter $\eta$ is defined conventionally as for a two site system (aka the Spin-Boson model). The definition via $-\eta \dot{x}$ is not practical because $x$ is a discrete variable. Still, in a semiclassical context, disregarding an ambiguity regarding numerical prefactor, the dissipation parameter $\eta$ has the meaning of a friction coefficient, as for X-coupling. With our standard conventions we get $C_{\parallel}{=}1/2$ (for S-dissipation) instead of $C_{\parallel}{=}1$ (for X-dissipation), which reflects a convention and not a profound difference. In contrast the $A$ coefficients are independent of convention and reflect quantum signature.

\subsection{B-dissipation}

For completeness we also consider the case where the dissipation is  
due to uncorrelated noisy bonds (rather than sites). 
Here we have an additional term in the expression 
for the diffusion coefficient, namely ${D =  D_{\parallel} + D_{\perp} }$, 
where 
\beq \label{eDXB}
D_{\perp} \ \approx \ C_{\perp}  \left[1 + A_{\perp} \left( \dfrac{c}{T} \right)^2 \right] \nu_B 
\eeq
with ${ A_{\perp} = -1/4 }$. This additional term reflects extra spreading in the space 
due to stochastic hopping as discussed in previous publication \cite{qss_sr}.
The quantum fingerprints are not related to this term, 
but to the $D_{\parallel}$ term that arises due to the noise-induced 
spreading in~$p$. For that term we find $A_{\parallel}{=}0$.

\sect{Technical note:} 
The dissipation parameter $\eta_B$ is defined as for the Spin-Boson model. It is the ``same" Ohmic bath as assumed for S-dissipation, but the coupling term is different. We get $C_{\parallel}{=}1/6$ and ${C_{\perp}{=}1}$, 
as opposed to X/S-coupling for which ${C_{\perp}{=}0}$.

\subsection{Non-universality}

In general, for high temperatures, the diffusion is composed of a term that originates 
from non-coherent hopping that is induced by the bath, namely $D_{\perp}$ of \Eq{eDXB}, 
and from the interplay of noise with coherent hopping between sites, namely $D_{\parallel}$ of \Eq{eDXS}.
The $\ell$ dependence of the latter is a quantum signature, and consequently  
our result for $D_{\parallel}$ reflect details of the dissipation mechanism.   

The $A$ coefficients are {\em not} a matter of convention.
Rather they reflect the thermalization and the spreading mechanism, 
and hence indicate {\em quantum} manifestation.
We summarize our results:      
\beq \label{eA}
&& A_{\parallel} \ = \ \left\{ 
\amatrix{             
-5/16 & \ \text{for X-coupling} \cr
+1/16 & \ \text{for S-coupling} \cr
0   & \ \text{for B-coupling} 
} \right.
\eeq

Contrary to the X-coupling case, for local (S/B) dissipators 
the canonical $\rho$ is not the exact steady-state,
and satisfies  $\mathcal{L} \rho \sim O(\beta^3)$ rather than zero.
We shall explain that if we ad-hock correct the transition rates 
to get agreement with Boltzmann, 
the results for the $A$-s are modified as follows:
\beq \label{eAsc}
&& A_{\parallel} \ = \ \left\{ 
\amatrix{             
-1/32 & \ \text{for S-coupling} \cr
-1/16   & \ \text{for B-coupling} 
} \right.
\eeq

We emphasize again that the value of $A$ is a sensitive probe 
that is affected by the line-shape of the spreading kernel. 
Therefore its precise value is non-universal but depends on 
the weights of the quantum transitions. 
For completeness we introduce in \Sec{sec:stochastic} 
results for intermediate values of~$\ell$, 
demonstrating the crossover from S-coupling ($\ell{\sim}a$) to X-coupling ($\ell{\sim}\infty$).

\section{The Ohmic dissipator}
\label{sec:ohmic-dis}

The isolated chain is defined by the~$\bm{H}^{(c)}$ Hamiltonian. 
The X-dissipation scheme involves a single bath, 
with interaction term ${-\bm{W} F}$, 
where $\bm{W}$ is the position operator $\bm{x}$, 
and $F$ is a bath operator that induces Ohmic fluctuations 
with intensity $\nu$.     
More generally we assume a disordered thermal environment 
that is composed of numerous uncorrelated baths 
such that the interaction term is $\sum_{\alpha} \bm{W}_{\alpha} F_{\alpha}$, 
where $\alpha$ labels different locations. 
For S-dissipation ${\bm{W}_{\alpha} = \kb{x_{\alpha}}{x_{\alpha}}}$, 
leading to a fluctuating potential that dephases the different sites.   
For B-dissipation ${\bm{W}_{\alpha} = \kb{x_{\alpha}{+}1}{x_{\alpha}} + \text{h.c.}}$, 
which induces incoherent hopping between neighbouring sites. 
The Ohmic dissipator $\mathcal{L}^{(\text{X/S/B})} \rho$ takes the form  
\cite{qss_sr,cohen2012lecture}:
\beq \label{e2}
-\sum_{\alpha} \left(
\dfrac{\nu}{2} [\bm{W}_{\alpha}, [\bm{W}_{\alpha}, \rho]] 
+ \dfrac{\eta}{2}\, i [\bm{W}_{\alpha}, \{\bm{V}_{\alpha}, \rho\}]  
\right) \ \ 
\eeq
where ${\eta=\nu/(2T)}$ is the friction coefficient, and  
\beq \label{eFR}
\bm{V}_{\alpha} \ \equiv \ i[\bm{H}^{(c)}, \bm{W}_{\alpha}]
\eeq 
The friction terms represent the response of the bath 
to the rate of change of the $\bm{W}_{\alpha}$.
For X-dissipation ${\bm{V} = c\sin(\bm{p}) }$ is the velocity operator. 
If we treat the the friction term of \Eq{e2} in a semi-classical way, 
the expression for the dissipator in the Wigner phase-space representation $\rho_w(R,P)$ 
takes the familiar Fokker-Plank (FP) form with ${v=c\sin(P)}$, namely, 
\beq \label{eFP}
\mathcal{L}^{\text{FP}}\rho_w \ \ = \ \ \frac{\nu}{2}\partial_P^2 [\rho_w] - \partial_P [(f_0-\eta v) \rho_w ] 
\eeq
which is a sum of momentum-diffusion and momentum-drift terms. 
For the sake of later reference we have added to the friction force ($-\eta v$) a constant field~$f_0$. 

The X-dissipator leads to canonical steady-state for any friction and for any temperature. This is not the case for S/B-dissipation, for which the agreement of the steady-state with the canonical result is guaranteed only to second order in~$\eta$.  The reason for that is related to the proper identification of the ``small parameter" that controls the deviation from canonical thermalization. 
The X-dissipator induces transitions between neighboring momenta, 
and therefore the small parameter is $\Delta/T$, where the level spacing $\Delta$ goes to zero in the $L\rightarrow\infty$ limit, where $L$ is the length of the chain. But for local baths, the coupling is to local scatterers,
that create transitions to all the levels within the band.
Therefore the small parameter is~$c/T$, and canonical thermalization is expected only for~${c/T < 1}$.

\section{Semiclassical analysis for X-dissipation}
\label{sec:semi-X}

We shall argue later that for X-dissipation the semiclassical dynamics 
that is generated by $\mathcal{L}^{\text{FP}}$ is {\em exact} for the purposed 
of the~$A$ coefficient evaluation. 
Here we present the semiclassical solution.

In \Sec{sec:CL-classical} below we find the steady-state momentum distribution
in the presence of a constant field~$f_0$.  
In \Sec{sec:transport-coef}  we obtain for weak field ${\braket{v} = \mu f_0}$, 
where $\mu$ is the mobility. 
Then the diffusion coefficient is deduced from the Einstein relation, 
namely, ${D = \mu T}$, leading to \Eq{eq:D-cl}. 

Optionally we can calculate directly the velocity-velocity correlation 
function $\braket{v(t) v(0)}$ in the absence of an external field. 
This requires a rather complicated recursive procedure, see \App{sec:sine-corr}. 
The diffusion coefficient is obtained via
\beq \label{eq:D-vvcorr}
D \ \ = \ \ \int_{0}^{\infty} dt \, \braket{v(t) v(0)}
\eeq
The same result is obtained, namely \Eq{eq:D-cl}. 
Later we shall calculate the diffusion in a proper 
quantum calculation, this again yields the same result. See \Sec{sec:quantum-anal}.

\subsection{The steady-state}
\label{sec:CL-classical}

We consider a Brownian particle that is described by \Eq{eq:H-tb-1}, 
under the influence of thermal non-disordered fluctuating field (X-coupling).
Below we set ${a=1}$ for the lattice constant.  
A fully-quantum treatment of this model has been introduced by \cite{aslangul1986quantum,AslangulPeriodicPotential1987}, 
with focus on low temperature QBM regime, 
while here we focus on the high temperature regime.    
The semiclassical equations of motion are formally obtained by the 
substitution ${f_0 \mapsto F(t)}$, 
where the total force ${ F(t) =  f_0 + f(t) -\eta \dot{x} }$  
includes a stochastic term that 
has zero average with correlation function  $\avg{f(t)f(t')} = \nu \delta(t-t')$,
and an associated friction term with coefficient $\eta$, 
in additional to the bias term $f_0$. 
Thus we get the Langevin equation 
\beq
\dot{x} &=& \dfrac{\partial H}{\partial p}  \ =  \ c \sin{(p)}  \label{eq:x-dot}\\
\dot{p} &=& -\dfrac{\partial H}{\partial x} \ = \ f_0  - \eta \dot{x} + f(t)  \label{eq:p-dot}
\eeq
The steady-state for $p$ is solved by inserting \Eq{eq:x-dot} to \Eq{eq:p-dot}.
Changing notation  $p\mapsto\varphi$, 
and ${u(\varphi)=f_0-\eta c \sin(\varphi)}$,
and $D_{\varphi}=(1/2)\nu$,  
one get the equation ${\dot{\varphi} = u(\varphi) + f(t)}$, 
with the associated Fokker-Planck equation
\begin{align}\label{eq:fp-phi}
\dfrac{\partial}{\partial t} \rho(\varphi,t) = - \dfrac{\partial}{\partial \varphi} I, 
\end{align}
with 
\beq  \nonumber
I &=&  u(\varphi) \rho - D_{\varphi} \dfrac{\partial \rho}{\partial\varphi} 
\equiv  - D_{\varphi} \left[ V'(\varphi) \rho + \dfrac{\partial \rho }{\partial \varphi}  \right] 
\\ \label{eq:p-current}
&=& - D_{\varphi} e^{-V(\varphi)}  \dfrac{\partial}{\partial \varphi} \left[ e^{V(\varphi)} \rho \right]
\eeq
This equation describes motion in a tilted potential
\beq \nonumber
V(\varphi) \ &=& \ -\frac{\eta c}{D_{\varphi}} \cos(\varphi) - \frac{f_0}{D_{\varphi}} \varphi 
\\  \label{eq:v-phi}
\ \ &\equiv& \ \ W(\varphi) - \mathcal{E} \varphi
\eeq
The non-equilibrium steady-state (NESS) solution is 
\begin{align} \label{eq:rho-steady-state-w-I}
\rho(\varphi) \ \  = \ \ \left[ C - I \int_0^{\varphi} \frac{ e^{V(\varphi')} }{D_{\varphi}} d\varphi' \right] e^{-V(\varphi)} 
\end{align}
where the integration constant $C$ is determined 
by the periodic boundary conditions ${\rho_0(0) = \rho_0(2\pi)}$,
namely,
\beq
C \ \ = \ \ \frac{I}{1-e^{-2\pi\mathcal{E}}} \int_0^{2\pi} \frac{ e^{V(\varphi')} }{D_{\varphi}} d\varphi'
\eeq 
Simplifying, the final expression for the NESS is 
\beq \label{eq:rho-ss}
\rho(\varphi)  = \frac{I}{1{-}e^{-2\pi\mathcal{E}}} 
\left[\int_0^{2\pi} \frac{dr}{D_{\varphi}} e^{W(\varphi+r) - \mathcal{E} r}  \right] 
e^{-W(\varphi)}  \ \ \ \ 
\eeq
Where the  $\varphi$-current $I$ is determined by normalization.

\subsection{The transport coefficients}
\label{sec:transport-coef}

Reverting to the original notations the first order result in $f_0$ is  
${I = [2 \pi \mathrm{I}_0^2 (c/T)]^{-1} f_0}$, 
where $\mathrm{I}_n(x)$ is the modified Bessel function.
For zero field the canonical distribution is recovered:
\beq
\rho(p) \ \propto \ \exp[-W(p)] \ = \ \exp[(c/T) \cos{(p)}]
\eeq
Averaging over \Eq{eq:p-dot}, and using $\avg{\dot{p}} = 2 \pi I$, one obtains
\beq \nonumber 
\avg{\dot{x}} = \left[1  - 2 \pi I \right] \frac{f_0}{\eta} 
= \left[ 1 -  \mathrm{I}_{0}^{-2} \Big(\dfrac{c}{T} \Big) \right] \frac{f_0}{\eta} 
\ \equiv \ \mu f_0  \ \ \ \ \ 
\eeq
where $\mu$ is the so-called linear mobility. 
This result for $\mu$ is consistent with direct calculation of $D$
in accordance with the Einstein relation, namely ${\mu = D/T}$.
The \textit{direct} calculation of $D$ is more involved. 
It is obtained by calculating the variance of~$x$, after time~$t$, 
for a particle initially located at $x{=}0$:
\beq \nonumber
\avg{x^2} =  c^2 \int_{0}^{t} \int_0^t dt'dt''  \avg{\sin(\varphi_{t'})\sin{(\varphi_{t''})}}
\ \equiv \ 2 D t \ \ \ \ 
\eeq
Defining $S_1$ as the area of the sine correlation function 
we write $D = c^2 S_1$. The calculation of $S_1$ is outlined in \App{sec:sine-corr}.

\section{Semiclassical analysis for S-dissipation}
\label{sec:semi-S}

In the semiclassical treatment $x$ is regarded as as a continuous coordinate, 
and therefore we write 
\beq
\bm{W}_{\alpha} \ = \ u_{\alpha}(\bm{x}) \ = \ u(\bm{x}{-}x_{\alpha}) 
\eeq
that involves a short-range interaction potential ${u(r)}$. 
The fluctuating potential is  
\beq
\mathcal{U}(x,t) \ = \ \sum_{\alpha} F_{\alpha}(t) u(x{-}x_{\alpha})
\eeq
In the semiclassical analysis we define $\nu$ as the variance of ${f= -\mathcal{U}'(x,t)}$. 
These fluctuations have the same intensity at any~$x$ because we assume 
that the $x_{\alpha}$ are homogeneously distributed.
It follows automatically that $\eta=\nu/2T$ is the friction coefficient, 
as in the case of X-dissipation. See \cite{Cohen1997} for details. 
So in the semiclassical description we get the same Langevine equation,  
irrespective of the correlation distance $\ell$ that is determined by the width of $u(r)$.  

In the tight-binding {\em quantum} model, 
we define $\nu_S$ as the variance of the on-site fluctuation of the potential. 
With that we associate a fluctuating force intensity  
\beq
\nu \ \ = \ \ \frac{1}{\ell^2} \nu_{S} 
\eeq
where $\ell$ is the correlation scale. 
We set ${\ell \sim a}$  where $a{=}1$ is the lattice constant.
Consequently $\nu$, up to numerical factor, is the same as $\nu_S$. 
The price for having a vague definition for $\nu$ 
is the prefactor $C$ that we get in the formula for $D$.
This prefactor reflects that the semiclassical limit has an 
inherent numerical ambiguity due to the residual freedom 
in the choice of~$u(r)$.

\section{Semiclassical analysis for B-dissipation}
\label{sec:semi-B}

Using the same prescription as for the S-dissipation case, 
and ignoring commutation issues, we write 
${\sum_{\alpha} \left( \kb{x_{\alpha}{+}1}{x_{\alpha}} + \text{h.c.} \right)}$ 
as ${ [2 \cos{(\bm p)}] \kb{\bm x}{\bm x}}$, 
and get for the B-coupling term 
\beq
\bm{W}_{\alpha} \ = 
 \ [2 \cos{(\bm{p})}] \, u_{\alpha}(\bm{x})
\eeq
This means that motion with momentum ${ |p| \sim \pi/2}$ is not affected by the baths. 
This is an artifact of the semiclassical treatment, 
and does not hold for the quantum dynamics. 
Still, the semiclassical perspective provides some insight 
that helps to clarify how \Eq{eDXB} comes out.

The equations of motion that are derived from the full Hamiltonian 
are of Langevin-type:  
\beq \label{eLEQx}
\dot{x} &=& \left[ c + 2 \sum_{\alpha} u_{\alpha}(x) F_{\alpha}(t) \right] \sin{(p)}  
\\ \label{eLEQp}
\dot{p} &=& \left[  2\sum_{\alpha} u'_{\alpha}(x)  F_{\alpha}(t) \right] \cos{(p)}
\eeq
For infinite temperature the $F_{\alpha}$ are uncorrelated white noise terms,  
with some intensity proportional to~$\nu_B$. 
Therefore we get from \Eq{eLEQp} diffusion in~$p$ 
with coefficient ${ \nu_p = (1/\ell)^2 [2\cos(ap)]^2 \nu_B }$,  
and from \Eq{eLEQx} extra diffusion in~$x$
with coefficient ${ \nu_x = (a)^2 [2\cos(ap)]^2 \nu_B }$,  
where ${\ell \approx a}$ and $a{=}1$. 
The latter term, after momentum averaging, is responsible for getting
the $D_{\perp}$ term in \Eq{eDXB}.
For a particle that moves with constant momentum ${p}$, ignoring the variation in $p$, the velocity-velocity correlation decays as ${\exp(-\nu_x t)}$
due to this $x$-diffusion. This leads to an extra Drude term ${D_{\parallel} = v^2 / \nu_x}$ that diverges at $p{=}\pi/2$. However, taking the variation of the momentum into account, this divergence has zero measure, and the final result is finite, leading to the first term in \Eq{eDXS} with ${C_{\parallel} = 0.49 }$. 
For finite temperature the fluctuations gain a non-zero average 
${\avg{F_{\alpha}} = 2 \eta_B \left([u_{\alpha}(x) \sin{(p)}] \dot{p} - [u'_{\alpha}(x)\cos{(p)}] \dot{x}  \right)}$, where ${\eta_B = \nu_B / T}$, 
leading to canonical-like thermalization, and over-estimated ${A_{\parallel} = -0.2}$. 
The results for $A_{\parallel}$ and $C_{\parallel}$ were obtained using a procedure  
that is described in the \Sec{sec:stochastic},
where we treat the {\em quantum} and the {\em semiclassical} on equal footing:
the latter can be regarded as a special case of the former.

\section{The quantum analysis}
\label{sec:quantum-anal}

The quantum evolution is generated by $\mathcal{L}$ of \Eq{e1} with the dissipators of \Eq{e2}, 
and it can be written as sum of Hamiltonian, noise and friction terms, 
namely ${\mathcal{L} = c\mathcal{L}^{(c)} + \nu \mathcal{L}^{(\nu)} + \eta c \mathcal{L}^{(\eta)}}$. 
various representations can be used, notably the Wigner and the Bloch representations see \App{sec:wigner}.
For the purpose of finding the spectrum (and from that the transport coefficients) 
it is most convenient to use the latter (Bloch), as explained below.

The elements of the super-vector $\rho$ are given in the standard representation 
by ${\rho(R,r) \equiv \BraKet{R+r/2}{\rho}{R-r/2}}$,
and in Dirac notation we write  $\rho = \sum_{R,r} \rho(R,r) \ket{R,r}$.
The super-matrix $\mathcal{L}$ is invariant under $R$-translations, 
and therefore it is convenient to switch to a Bloch representation $\rho(q;r)$ 
where $\mathcal{L}$ decomposes into $q$~blocks. In the $q$ subspace we have 
the following expressions \App{sec:bloch}:
\beq 
\label{eq:L-H-bloch-nongauged} \nonumber
\mathcal{L}^{(c)} &=& 
+\sin(q/2) \Big(\mathcal{D}_{\perp} - \mathcal{D}_{\perp}^{\dag} \Big)
\\
\label{eq:L-X-bloch} \nonumber
\mathcal{L}^{(\nu_X)} &=& - (1/2)\hat{r}^2 \\ \nonumber
\mathcal{L}^{(\eta_X)} &=&   \cos{(q/2)} \dfrac{\hat{r}}{2} \left( D_{\perp} -  D_{\perp}^{\dag} \right) 
\\
\label{eq:L-S-bloch} \nonumber
\mathcal{L}^{(\nu_S)} &=& -1 + 1 \kb{0}{0} \\ \nonumber
\mathcal{L}^{(\eta_S)} &=& 
\dfrac{\cos{(q/2)}}{2} \Big( 
\mathcal{D}_{\perp} + D_{\perp}^{\dag} + \kb{\pm 1}{0} - \kb{0}{\pm 1} \Big)
\\
\nonumber
\mathcal{L}^{(\nu_B)} &=& -2 \ + \ 2\cos(q) \kb{0}{0} 
+ \Big(\kb{1}{{-1}} + \kb{{-1}}{1} \Big) \\ 
\nonumber
\mathcal{L}^{(\eta_B)} &=&  \frac{1}{2}\cos{(q/2)} \Big(\mathcal{D}_{\perp}+\mathcal{D}_{\perp}^{\dag} \Big)  \\ \nonumber
&+& \dfrac{1}{2} \cos(3q/2) \Big(  \kb{\pm 1}{0} - \kb{0}{\pm 1} \Big) \\ 
&+& \dfrac{1}{2}\cos(q/2) \Big( \kb{{\mp 2}}{\pm 1} - \kb{\pm 1}{{\mp 2}} \Big)
\label{eq:L-B-bloch} \label{eLterms}
\eeq
The subscripts X/S/B distinguish the different coupling schemes, and $\mathcal{D}_{\perp} = \kb{r{+}1}{r}$ is the displacement operator in~$r$ space.

\subsection{Extracting the diffusion coefficient}

To obtain the diffusion coefficient, we consider the spectrum of $\mathcal{L}$ for a finite system of $L$ sites. In the Bloch representation the equation ${\mathcal{L} \rho = - \lambda \rho}$ decomposes into $q$-blocks. For a given~$q$ we have a tight binding equation in the $\ket{r}$ basis. For example $\mathcal{L}^{(c)}$ induces near-neighbor hopping in~$r$.  
The eigenvalues for a given $q$ are labeled $\lambda_{q,s}$, where ${s}$ is a band index. The long-time dynamics is determined by the slow ($s{=}0$) modes. Specifically, the diffusion coefficient is determined by the small~$q$ expansion 
\beq \label{elambda}
\lambda_{q,0} \ \ = \ \ D q^2 + \mathcal{O}(q^4)  
\eeq  
The NESS eigenvector belongs to the $q{=}0$ block, and for $\eta{=}0$ it is given by $\ket{r{=}0}$.
Non-zero $q$ and $\eta$ can be treated as a perturbation. The key observation is that in order to get an {\em exact} result for $D$ it is enough to use second-order perturbation theory in~$q$. The outcome of this procedure is the analytical expression for $D$ with the associated results for the $A$ coefficients. 
Extra technical details are provided in the next subsection.

\subsection{Perturbation theory}
\label{sec:perturbation}
 
We use perturbation theory to find the eigenvalue $\lambda_{q,0}$ of $\mathcal{L}^{(q)}$, from which we can obtain~$D$. We regard the Bloch quasimomentum~$q$ and the friction~$\eta$ as the perturbation. For ${q=\eta=0}$ the state $\ket{r={0}}$ is an exact eigenstate that is associated with the eigenvalue ${\lambda=0}$. Due to the perturbation it is mixed with neighboring $\ket{r}$ states. We outline below how we get analytical expressions for $\lambda_{q,0}$ to any order in $q$ and $\eta$. In practice we go up to second order. 

In the following we demonstrate how we perform perturbation theory for the X-coupling scheme. The same method is
used for the S/B coupling schemes either with the Ohmic dissipators or with the Boltzmann dissipators. 
We would like to diagonalize the $q$~block  
\beq \nonumber
\mathcal{L}^{(q)} & = & c \mathcal{L}^{(c)} + \nu \mathcal{L}^{(\nu_X)} + (c \eta)\mathcal{L}^{(\eta_X)} 
\\ \nonumber
&=& c \sin(q/2) \Big(\mathcal{D}_{\perp} - \mathcal{D}_{\perp}^{\dag} \Big) - (\nu/2)\hat{r}^2  
\\ \label{eq:L-X-SM}
&& + (c \eta) \cos{(q/2)} \dfrac{\hat{r}}{2} \left( D_{\perp} -  D_{\perp}^{\dag} \right)
\eeq
Each such block produces eigenvalues $\mathcal{L}^{(q)} \ket{s} = -\lambda_{q,s} \ket{s}$,
that are distinguished by the index $s$. We are interested in the slowest mode $\lambda_{q,0}$.
The NESS is the eigenvector that corresponds to the zero eigenvalue.
It belongs to the $q{=}0$ block, which results from probability conservation.
In the Bloch representation, probability conservation means that ${\bra{0}\mathcal{L}^{(0)} =0}$.
To obtain the eigenvalues to order $q^2$ it is enough to Taylor expand the operator to that order.
Accordingly, 
\beq \nonumber
\mathcal{L}^{(q)} \ \ &=& \ \  
- (\nu/2)\hat{r}^2 
\ + \ c (q/2) \Big(\mathcal{D}_{\perp} - \mathcal{D}_{\perp}^{\dag} \Big) 
\\ \label{eq:L-X-taylor}
&& + \ (c \eta) \left[ 1 - (q/2)^{2} \right] \dfrac{\hat{r}}{2} \left( D_{\perp} -  D_{\perp}^{\dag} \right)
\eeq
The first term is the zero order term. Here (for X-coupling) it is diagonal in~$r$.
For the other coupling schemes it is not necessarily diagonal in~$r$, 
but for any of them $\ket{r={0}}$ is an eigenstate of the zero-order term.

To find the eigenvalue $\lambda_{q,0}$ via perturbation theory one has to sum over different paths that begin and end in $r{=}0$. In the case of \Eq{eq:L-X-taylor} these paths  are composed of hops between near neighbor sites. Second order contributions involve terms with  
$\bra{0} \mathcal{L}^{(q)} \kb{r}{r} \mathcal{L}^{(q)} \ket{0}$, with $r{\ne}0$.
Each transition involves a factor $cq$ or $(c \eta)$, or $(c \eta q^2)$.
Hence only the sites ${|r| \le 2}$ contribute to the perturbed eigenvalue up to order~$\eta^2q^2$.
Furthermore, the $(c \eta q^2)$ transitions are always multiplied by other $\mathcal{O}(q)$ transitions, 
and therefore can be ignored in any second order expansion. 

From the above it should be clear that for X-coupling the matrix that should be diagonalized is
\beq \nonumber
\mathcal{L}^{(q)} \mapsto  
 \dfrac{1}{2} 
\left(
\begin{array}{ccccc}
 -4 \nu &  2 c \eta  {-} cq   & 0 & 0 & 0 \\
  -c \eta  {+}  c q & -\nu &   c \eta  {-} c q  & 0 & 0 \\
 0 & c q & 0 & -c q & 0 \\
 0 & 0 &  c \eta {+} c q   & - \nu &  -c \eta {-} c q  \\
 0 & 0 & 0 & 2 c \eta {+} c q  & -4 \nu \\
\end{array}
\right)
\eeq  
A convenient way to obtain analytical result is to write the characteristic equation ${\det[\lambda + \mathcal{L}^{(q)}] = 0}$ with the above (truncated) matrix, and to substitute an expansion ${\lambda_{q,0} = \sum_{n} a_n q^n}$. Then we solve for the coefficients $a_n$ iteratively. The outcome is expanded in $\eta$ to order $\eta^2$. Note that to go beyond second order in $\eta$ does not makes sense, because the Ohmic master equation and the associated NESS are valid only up to this order.

\section{Effective stochastic description}
\label{sec:stochastic}

The propagation of the Wigner distribution function $\rho_w(R,P)$ is generated by a kernel ${\mathcal{L}(R,P|R_0,P_0)}$  that is obtained from \Eq{eLterms} in a straightforward manner via Fourier transform \App{sec:wigner}. For simulations of the long time spreading it is enough to approximate $\mathcal{L}$ in a way that is consistent with second order perturbation theory in~$q$. As explained in the previous paragraph, such approximation provides an {\em exact} result as far as $D$ calculation is concerned.    
Replacing $\sin(q/2)$ by $(q/2)$, the $\mathcal{L}^{(c)}$ term by itself generates classical motion in the $X$ direction with velocity ${v=c\sin(P)}$. In the quantum calculation this motion is decorated by a Bessel function, but~$D$ is not affected. 
The $\cos(q)$ in the $\mathcal{L}^{(\nu_B)}$, after expansion to second order and Fourier transform, leads to an $x$-diffusion term, that is responsible to for the ${C_{\perp} }$ contribution in \Eq{eDXB}. As far as this term is concerned, there is no difference between the quantum and the semiclassical picture, and therefore we ignore it in the subsequent analysis.   
The cosine factors in the other dissipators can be replaced by unity. The reason is as follows: by themselves those cosine terms do not lead to any diffusion; only when combined with the  $\mathcal{L}^{(c)}$ term they lead to the Drude-type ${C_{\parallel} }$  contribution in \Eq{eDXS}; the $\mathcal{L}^{(c)}$ is already first order in~$q$; hence no need to expand the cosines beyond zero order.

\subsection{The effective rate equation}

With the approximations that were discussed in the previous paragraph (excluding for presentation purpose the trivial $R$ diffusion in the case of B-disspation), we find that the evolution of the Wigner function is generated by a stochastic-like kernel ${\mathcal{L}(R,P|R_0,P_0) = \mathcal{W}(P|P_0) \delta(R-R_0)}$. The explicit expressions for infinite temperature ($\eta{=}0$) are:   
\beq
\label{eW15}
\mathcal{W}^{(\nu_X)}(P|P_0) &=& \left(\frac{L}{2\pi}\right)^2 \dfrac{\nu}{2} \,  \delta_{P, P_0 \pm (2\pi/L)}   \\ 
\label{eW16}
\mathcal{W}^{(\nu_S)}(P|P_0) &=& \left(\frac{\nu_{S}}{L}\right)  \\
\label{eW17}
\mathcal{W}^{(\nu_B)}(P|P_0) &=& \left(\frac{\nu_{B}}{L}\right) 4\cos^2{ \left( \dfrac{P+P_0}{2}  \right)} 
\eeq
These are the transition rates (${P\ne P_0}$), while the diagonal elements of $\mathcal{W}$ are implied by conservation of probability. For X-dissipation \Eq{eW15} describes local spreading of momentum which is in complete correspondence with the semiclassical analysis.
The noise intensity is reflected in the second moment: 
\beq \label{eWrr}
\nu \ \  = \ \ \sum_{p} W(p) \, p^2  
\eeq
where ${p = (P - P_0)}$.
This implies consistency with the Langevin equation \Eq{eq:langevin-p}.
Optionally \Eq{eW15} can be regarded as the discrete version 
of the Fokker-plank equation \Eq{eFP}.
For S-dissipation \Eq{eW16}  describes quantum diffractive spreading. In the latter case, if the dynamics were treated semiclassically one would obtain the same result as for X-dissipation, namely \Eq{eW15}, with prefactor of order unity that can be by re-scaled to unity by adopting the appropriate convention for the definition of~$\nu$. In other words: the coupling strength to the bath should be re-defined such that $\nu$ is the second-moment of $\mathcal{W}(P|P_0)$ irrespective of the lineshape. Similarly, if  the dynamics were treated semiclassically for the B-coupling, one would obtain \Eq{eW15} multiplied by ${4\cos^2(P)}$, as implied by the semiclassical analysis.

The result for $\mathcal{W}$ for finite temperature, 
in leading order in $\eta$ (which serves here as a dimensionless version of the inverse temperature) can be written as  
\beq\label{eq:Wkk-Boltzmann}
\mathcal{W}(P|P_0) = \mathcal{W}^{(\nu)}(P|P_0)  
\exp\left[- \dfrac{E(P) {-} E(P_0)}{2 T}\right]
\ \ \ \ 
\eeq
where ${E(P)=-c \cos(P)}$. More precisely, if we incorporate the $\mathcal{L}^{(\eta)}$ term of the Ohmic master equation, we get \Eq{eq:Wkk-Boltzmann} with $e^x \mapsto (1+x)$. This reflects the well known observation that the Ohmic approximation satisfies detailed balance to second order in $\eta$.
Accordingly the Ohmic steady-state agrees to {\em second order} with the canonical steady-state ${\rho_{\text{SS}}(P) \propto \exp[-E(P)/T] }$.

\subsection{Analytical and numerical estimates}
\label{sec:simulation}

The stochastic description allows a convenient way to obtain {\em exact} results for~$D$ 
either analytically or numerically. 
Analytically we use the same procedure as in the quantum case, 
namely, given the dissipator $\mathcal{L}^{(q)}$, 
we extract $D$ from \Eq{elambda}. The relation between $\mathcal{L}^{(q)}$ and the 
stochastic kernel is 
\beq \nonumber
&& \BraKet{r}{\mathcal{L}^{(q)}}{r_0} \ = \ \BraKet{r,q}{\mathcal{L}}{r_0,q}
\\ \label{eLqW}
&& \ \ \ \ \ = \ \dfrac{1}{L} \sum_{P, P_0} \mathcal{W}(P|P_0) e^{i P r-i P_0 r_0}
\eeq

\begin{figure}
\centering
\includegraphics[width=\hsize]{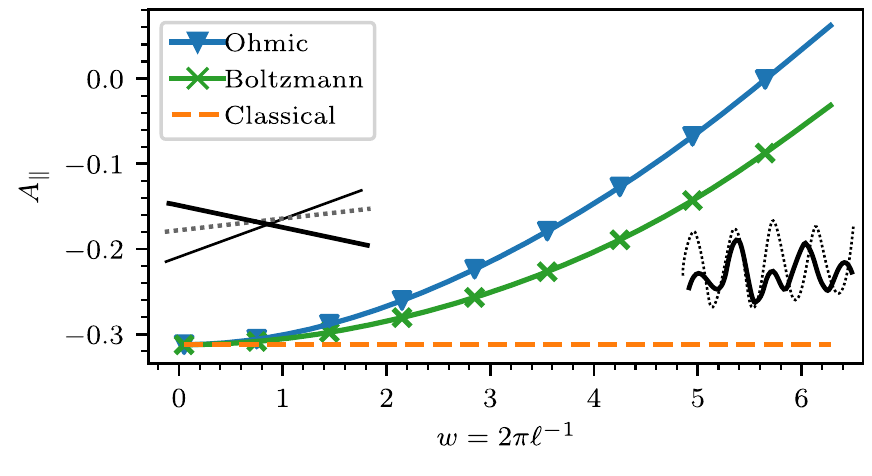}
\caption{\label{fig:a-vs-ellL} 
{\bf Quantum non-universality}.
The dependence of the coefficient $A_{\parallel}$ on $\ell$. 
The insets caricature the fluctuating potential 
for large (left) and for small (right) values of~$\ell$.
In the semiclassical analysis the result (dashed orange line) is universal, independent of~$\ell$.    
In the quantum analysis we obtain an interpolation between 
the X-coupling and the S-coupling case \Eq{eA}. 
We plot results (see text) for the Ohmic (blue triangles) 
and for the Boltzmann-corrected (green crosses) 
versions of the master equation.   
}
\end{figure}

The X-coupling and S-coupling schemes provide two extremes, with ${\ell = L}$ and ${\ell = 1}$ respectively.
This is mirrored in the infinite-temperature kernel $\mathcal{W}$ of \Eq{eW15} and \Eq{eW16}.
On equal footing we can interpolate between the two extremes by introducing a kernel 
of width $2 \pi/\ell$. Then we use \Eq{eq:Wkk-Boltzmann} to get the finite-temperature kernel. 
The calculation of  $\mathcal{L}^{(q)}$ using \Eq{eLqW} is provided in \App{sec:label}. 
The result for $A_{\parallel}$ is displayed in \Fig{fig:a-vs-ellL}. 
Note that the convention regarding the prefactor in $\mathcal{W}^{(\nu)}$ 
plays no role in the determination of $A_{\parallel}$.

At this point we have to emphasize again that for the ``Ohmic" results we 
use the prescription  ${e^x \mapsto (1+x)}$ as explained after \Eq{eq:Wkk-Boltzmann}.
If we perform the calculation literally using \Eq{eq:Wkk-Boltzmann} we get \Eq{eAsc} instead of \Eq{eA}.
Note that the same results are obtained with ${e^x \mapsto (1+x+(1/2)x^2)}$,
because higher orders do not affect the expansion in \Eq{elambda}.
The difference between \Eq{eAsc} and \Eq{eA} reflects the limited accuracy 
of the Ohmic master equation with respect to the small parameter~$c/T$.

The analytial results for the $A_{\parallel}$ coefficients that are plotted in \Fig{fig:a-vs-ellL}
are derived and displayed in \App{sec:label}. Here we write expressions that 
approximate very well the exact results:
\beq\label{eq:A-w-quad}
A_{\parallel} &\approx& -\dfrac{5}{16} \left( 1 - \dfrac{6}{5} \left( \dfrac{a}{\ell} \right)^2 \right)  
\ \ \ \ \ \mbox{[Ohmic]} \\
A_{\parallel} &\approx&  -\dfrac{5}{16} \left( 1 - \dfrac{9}{10} \left( \dfrac{a}{\ell} \right)^2\right) 
\ \ \ \ \ \mbox{[Boltzmann]}
\eeq
Note that this practical approximation provides the exact results  
for both X-coupling ($\ell{=}\infty$) and S-coupling ($\ell{=}a{=}1$).

In \Fig{fg1} we test the analytical approximation \Eq{eDXS} against exact numerical calculation that is based on the effective rate equation. In the numerical procedure the diffusion coefficient~$D$ is calculated using \Eq{eq:D-vvcorr}. The momentum spreading kernel is ${K(t) \equiv  \exp(\mathcal{W} t)}$, and the velocity is ${v_P = c \sin(P)}$. Accordingly 
\beq \label{eq:vvcorr}
\avg{v(t)v(0)} =  \sum_{P,P_0} v_{P} [K_{P,P_0}(t)] v_{P_0}  \rho_{\text{SS}}(P_0)
\eeq
If we perform the calculation literally using \Eq{eq:Wkk-Boltzmann} we get results that agree with \Eq{eAsc}. 
If on the other hand we use for $\mathcal{W}$ the Ohmic expression (as specified after \Eq{eq:Wkk-Boltzmann}) 
we get results that agree with \Eq{eA}. 
Note that for $\rho_{SS}$ we can use the canonical steady state, 
because it agree with the Ohmic steady state to second order.

\newpage
\section{Discussion}
\label{sec:discussion}

The prototype Caldeira-Leggett model corresponds to the standard Langevin equation where the dispersion relation is ${v=(1/\mass)p}$. In the tight-binding framework we have the identification ${ \mass \mapsto 1/(c a^2) }$, where $a$ is the lattice constant. There is a crossover to standard QBM as ${\theta \equiv T/c}$ is lowered. 
It is illuminating to summarize this crossover in terms of mobility.  
Using the Einstein relation \Eq{eDXS} and \Eq{eDXB} imply  
\beq \label{eMOB}
\mu \ \ = \ \ \frac{D}{T} \ \ = \ \ \frac{B(\theta)}{\eta} \ + \ 2 Q(\theta) \eta
\eeq
where the $B(\theta)$ term is related to the coherent hopping,
and the $Q(\theta)$ term is due to bath-induced incoherent hopping.
We believe that this functional form is rather robust, and apply to any type of dissipation mechanism.
The traditional result is the first term with ${B(\theta)=1}$,
while \Eq{eDXS} implies that for large~${\theta}$ the result is  
\beq
B(\theta) \ \propto \ (1/\theta)^2+A_{\parallel}(1/\theta)^{4} 
\eeq
We have shown how $A_{\parallel}$ depend on $\ell$, 
with emphasis on the extreme limits of X-coupling and S-coupling. We conclude that the $A$ coefficients provide a way to probe the underlying mechanism of dissipation, and to identify the {\em high-temperature fingerprints} of quantum mechanics.

It should be instructive to demonstrate experimentally that $\mu(T;\ell)$ indeed depends on~$\ell$. Ref.\cite{muMeas} provides an experimental demonstration of measuring mobility versus temperature for a semiconductor device, while Ref.\cite{KARL2003649} reviews experimental methods used to extract the mobility in organic semiconductors.
Consider the possibility of fabricating a metallic {\em gate} that produces thermal electrostatic fluctuations. Metals that differ in their {\em granularity} are characterized by a different form factor, with different correlation scale $\ell$.
Thus it would be possible to demonstrate that $\ell$ has significance. Hopefully it would be possible to further extract, experimentally, the non-universal dependence of $A_{\parallel}$ on the correlation distance $\ell$, and to test the prediction of \Fig{fig:a-vs-ellL}.

\ \\
\sect{Acknowledgment}
This research was supported by the Israel Science Foundation (Grant No.283/18). We thank Muntaser Naamneh for his advice on the experimental aspect.



\clearpage
\onecolumngrid
\pagestyle{empty}

\appendix

\section{The sine correlation function}
\label{sec:sine-corr}

First we recall that for zero field the steady-state is an equilibrium canonical state 
${\rho(\varphi) \propto \exp[-W(\varphi)]}$,  
where  ${W(\varphi) = z \cos(\varphi)}$, 
and ${z=(c/T)}$. At equilibrium we have     
\beq \label{ewn}
w_n \ \ \equiv \ \ \avg{\cos{(n \varphi)}} \ \ = \ \ \frac{\mathrm{I}_n(z)}{\mathrm{I}_0(z)}
\eeq
We define $S_n$ as the area of the sine-sine correlation function 
${s_n(t) =  \avg{\sin(n\,\varphi_t) \sin(\varphi_0)} }$, namely,
\beq
S_n  \ \ = \ \ \int_0^{\infty }  s_n(t) dt, 
\hspace{2cm} n=0,1,2,...
\eeq
Eventually we are interested only in $S_1$, but for the derivation we define a full set of sine correlation functions.
Explicitly these are written as
\begin{align}
s_n(t) \ \ 
\ \ = \ \ \int_0^{2\pi}  \avg{\sin(n\,\varphi_t)}_0  \sin{(\varphi_0)} \rho(\varphi_0) d\varphi_0
\ \ = \ \ \int_0^{2\pi}  \avg{\sin(n\,\varphi)}_t  \sin{(\varphi_0)}  \rho(\varphi_0) d\varphi_0
\end{align}
The average without subscript assumes equilibrium state, while the average with subscript ``0'' indicates initial condition~$\varphi_0$ and assumes a Langevin picture. The subscript ``t'' indicates expectation value after time~$t$ within the framework of the associated Fokker-Planck picture. Initially we have 
\begin{align}\label{eq:sn-bessel}
s_n(0) \ \ = \ \ \avg{\sin{(n \varphi)} \sin{(\varphi)}} 
\ \ = \ \ \dfrac{1}{2} \dfrac{ \mathrm{I}_{n-1}(z)  - \mathrm{I}_{n+1}(z) }{\mathrm{I}_0(z)} 
\ \ = \ \  \dfrac{n}{z} \dfrac{\mathrm{I}_{n}(z)}{\mathrm{I}_0(z)}
\end{align}
In order to find $s_n(t)$ at later times, we realize that the it satisfies the same equation of motion as that of $\avg{\sin(n \varphi_t)}_{0}$, where $0$ indicates any initial state. This is known as the ``regression theorem''. The adjoint equation for any observable ${A(\varphi)}$ is  
\begin{align}\label{eq:fp-adjoint}
\dfrac{\partial}{\partial t} \braket{A(\varphi)}_t  \ \ = \ \ 
\avg{ 
D_\varphi \left(\dfrac{\partial^2}{\partial{\varphi^2}} - {W'(\varphi)} \dfrac{\partial}{\partial{\varphi}} \right) A(\varphi) }_t 
\end{align}
Substituting ${A(\varphi) := \sin(\varphi)\sin(\varphi_0) }$,  
and integrating over time, one obtains a recursive equation for the $S_n$, 
\beq \label{esn}
s_n(0) \ = \ n^2 \frac{\nu}{2} S_n + n \dfrac{\eta c}{2} \left(S_{n+1} - S_{n-1} \right) 
\eeq
with the boundary conditions ${S_0 = S_{\infty } = 0}$. 
At this point it is useful to realize that from \Eq{eq:fp-adjoint} with ${A(\varphi) := \cos(n \varphi) }$ 
it follows that the {\em stationary} values $w_n$ of \Eq{ewn} obey \Eq{esn} with zero on the left hand side.   
It is therefore useful to substitute $S_n := w_n \tilde{S}_n $ 
in order to get a first order difference equation for the $\tilde{S}_n $
that can be solved by recursion. 
The procedure is explained with details in Section VII of 
[\href{https://link.aps.org/doi/10.1103/PhysRevE.96.042152}{Shapira and Cohen, Phys. Rev. E 96, 042152 (2017)}]
and leads to the solution
\begin{align}
S_1  \ = \ 
- \dfrac{1}{\eta c}\sum_{n=1}^{\infty} \dfrac{(-1)^{n}}{n} s_n(0) w_n 
\ = \ -\dfrac{\nu}{(\eta c)^2} \sum_{n=1}^{\infty} (-1)^n \left[ \dfrac{\mathrm{I}_n(z)}{\mathrm{I}_0(z)}  \right]^2  
\ = \ \dfrac{\nu}{2 (\eta c)^2} \left[1 -  \mathrm{I}^{-2}_{0}(z)\right]
\end{align}
Where we used the completeness relation 
\begin{align}\label{eq:bessel-summation}
1 \ \ = \ \ \mathrm{I}_0^2(z) + 2 \sum_{n}  \mathrm{I}_n^2(z)(-1)^n
\end{align}

\clearpage

\section{The Wigner phase space representation}
\label{sec:wigner}

Here we treat ${(x,p)}$ as extended continuous coordinates and derive the standard Wigner representation for the quantum propagation in the absence of dissipators. The elements of the  $\rho$ are given in the standard space representation 
by ${\rho_{x',x''} \equiv \BraKet{x'}{\rho}{x''}}$. We define ${r=x'-x''}$ and ${R=(x'+x'')/2}$, and use super-vector Dirac notations, namely  $\rho = \sum_{R,r} \rho(R,r) \ket{R,r}$.
The space representation is $\rho(R,r) \equiv \rho_{x',x''}$, the momentum representation $\rho(q,P)$ is related by double Fourier transforms, 
and the intermediate representations are those of Wigner $\rho_w(R,P)$ and Bloch $\rho(q;r)$. 
For the unitary evolution with ${ U=\exp[i ct \cos(\bm{p})] }$, the propagator of the Wigner function in momentum representation is 
\beq
\mathcal{K}(q,P|q_0,P_0) \ &=& \ \BraKet{P{+}(q/2)}{U}{P_0{+}(q_0/2)} \, \BraKet{P{-}(q/2)}{U}{P_0{-}(q_0/2)}^*  \\
\ &=& \  2\pi \delta(q-q_0) \ 2\pi \delta(P-P_0) \exp\left[-i 2ct \sin(q/2) \sin(P) \right]
\eeq 
leading to 
\beq
\mathcal{K}(R,P|R_0,P_0) \ = \ 2\pi \delta(P-P_0) \int \frac{dq}{2\pi}  \exp\left[-i 2ct \sin(q/2) \sin(P)  + iq (R-R_0)\right] 
\eeq
Note that this kernel is properly normalized with respect to the integration measure $dRdP/(2\pi)$. \\
With ${\sin(q/2) \mapsto (q/2)}$ we get the classical result 
\beq
\mathcal{K}(R,P|R_0,P_0) \ \ = \ \ 2\pi \delta(P-P_0) \ \delta((R-R_0) - ct \sin(P))
\eeq
But quantum mechanically we get 
\beq
\mathcal{K}(R,P|R_0,P_0) \ = \  \sum_n 2\pi \delta(P-P_0) \delta((R-R_0) - n)  \bm{J}_{2n} \left( 2ct \sin(P) \right) 
\eeq
In the above sum $n$ runs formally over all the integer and half-integer values. Note that Wigner function on a lattice has support on both integer and half integer lattice points (weight on half integer lattice points is the fingerprint of interference due to superposition of integer lattice locations).

\section{The Bloch representation}
\label{sec:bloch}

For an infinite chain the conventional way to define the Bloch representation is to perform ${R \mapsto q }$  Fourier  transform of $\rho(R,r)$ for a given $r$ to obtain $\rho(q;r)$.  Note that $R$ runs over integer values for $r=0,2,4,...$  and over half integer values for $r=1,3,5...$. This definition has a problem if we consider a finite chain with periodic boundary conditions. Still it can be justified after a short transient if $L$  is large enough because distant points in space loose  phase correlation (if there was to begin with). For a small ring (small $L$) this might not be the case. Therefore in a previous work \cite{qss_sr} we have defined ad-hock the Bloch representation $\rho_q(r)$ as the Fourier transform of $\BraKet{x}{\rho}{x+r}$. The ad-hock definition differs by gauge transformation (and non-intentionally also by sign) from the conventional definition, and allows to handle correctly the periodicity in both coordinates, namely, also in $r$.    
For a small chain, or for a complete investigation of the eigenvalues problem, these phases are important. 
See for example \cite{EspositoGaspard2005}.

Our system is invariant under translations, therefore it is natural 
to perform the diagonalization of $\mathcal{L}$ is the Bloch representation.
In practice one can obtain the expressions in \Eq{eLterms} by inspection. 
As an example let us see how the expression for $\mathcal{L}^{(c)}$ is obtained. 
It originates from ${i[\cos(\bm{p}),\rho]}$. In the standard representation its matrix elements are
\beq
\mathcal{L}^{(c)}(x', x''|x'_0,x''_0)
\ = \ i\BraKet{x'}{ \cos(\bm{p}) }{x'_0} \delta(x''-x''_0)   
- i\delta(x'-x'_0)  \BraKet{x''}{ \cos(\bm{p}) }{x''_0}  
\eeq
Recall that ${ \cos(\bm{p}) }$ is the sum of displacement operators $e^{\mp i\bm{p}}$.   
In super-vector notations the above expression can be written in terms of  
operators $e^{\mp i (1/2)\bm{q}}$ and $e^{\mp i\bm{P}}$ that induce
translations in $R$ and in $r$ respectively. Namely,
\beq
\mathcal{L}^{(c)}(R, r|R_0,r_0)
\ = \ i\BraKet{R,r}{  \cos\left(\frac{\bm{q}}{2} +\bm{P} \right)}{R_0,r_0}  
- i\BraKet{R,r}{  \cos\left(\frac{\bm{q}}{2} -\bm{P} \right)}{R_0,r_0}
\eeq
Thus we can write
\beq
\mathcal{L}^{(c)}
\ = \ -i 2 \sin\left(\frac{\bm{q}}{2}\right) \sin\left(\bm{P}\right)
\ = \  \sin\left(\frac{\bm{q}}{2}\right) \Big[\mathcal{D}_{\perp} - \mathcal{D}_{\perp}^{\dag} \Big]
\eeq
In the Bloch $(q,r)$ representation this super-operator becomes block diagonal in~$q$.

\clearpage
\section{From Bloch to Wigner and back}
\label{sec:label}

In the main text we present in \Eq{eLterms} the Bloch representation ${\mathcal{L}^{(q)}}$ of the dissipators. 
The transformation to the Wigner representation is essentially a Fourier transform:
\beq \label{eq:L-wigner}
\mathcal{L}(R,P|R_0,P_0) \ = \  \int \frac{dq}{2\pi}  e^{i q (R-R_0)} \iint  dr dr_0 e^{-irP + ir_0P_0} \BraKet{r}{\mathcal{L}^{(q)}}{r_0}
\eeq
Note that the inner integral transforms $\BraKet{r}{\mathcal{L}^{(q)}}{r_0}$ to the momentum representation $\BraKet{P}{\mathcal{L}^{(q)}}{P_0}$. Note also that ${\mathcal{W}(P|P_0) = \BraKet{P}{\mathcal{L}^{(q{=}0)}}{P_0}}$ are the Fermi Golden Rule (FGR) transition rates between momentum eigenstates. Commonly the FGR is considered as an approximation, while we have rigorously established that  ${\mathcal{W}(P|P_0)}$ can be used within an effective rate equation in order to evaluate the exact quantum result for~$D$.

In the main text we use a discrete momentum notation, 
such that ${\Braket{r}{P}= L^{-1/2}\exp(iP R)}$, etc. 
Consequently, in the discrete version of \Eq{eq:L-wigner}, 
the integrand of the $dr dr_0$ integral contains an extra $1/L$ factor. 
On the other hand for summation $\sum_P$ over momenta
the measure becomes $[L/(2\pi)]dP$.  
 

{\bf Transforming to Bloch.-- }
It is convenient to handle the calculations of the spectrum  
on equal footing for all the coupling schemes, 
for both Ohmic and Boltzmann versions of the dissipators. 
For this purpose we have to transform  \Eq{eq:Wkk-Boltzmann}
back from the Wigner representation to the Bloch representation using \Eq{eLqW}.  
Note that this equation does not depend on $q$, 
reflecting the $\delta(R-R_0)$ of the transitions.  
Making the distinction between the diagonal terms ($P=P_0)$ and the off-diagonal terms ($P \ne P_0$),
taking into account that by definition the kernel conserves probability, 
namely, ${\sum_P \mathcal{W}(P|P_0) = 0}$, one can write 
\beq 
\BraKet{r}{\mathcal{L}^{(q)}}{r_0}
\ = \ \tilde{\mathcal{W}}(r,r_0) - \tilde{\mathcal{W}}(0,r_0-r) 
\eeq
where
\beq \label{eLqWW}
\tilde{\mathcal{W}}(r,r_0) \ = \ 
\dfrac{1}{L}\iint  \mathcal{W}(P|P_0)  \, e^{i P r-i P_0 r_0}  \, \frac{L}{2\pi}dP \, \frac{L}{2\pi}dP_0 
\eeq 
In the latter expression it is implicit that $P=P_0$ has measure zero, 
so it reflects the contribution of the ${P\ne P_0}$ terms in the discrete sum of \Eq{eLqW}.  
Finally, note that for $\eta{=}0$ the Bloch kernel is diagonal 
(the non-zero elements are those with ${r=r_0}$), 
and that from \Eq{eWrr} it follows that for $r = r_0 = \pm 1$ 
we have by convention ${\BraKet{r}{\mathcal{L}^{(q)}}{r_0} = -(\nu/2) }$.

{\bf General kernel. -- }
We consider a kernel $\mathcal{W}(P|P_0)$ of width $w = 2\pi/\ell$.
Its normalization $C$ should be determined such that the $dp$ integral 
over ${[1-\cos(p)]}$ equals~$1/2$ (see the last sentence of the previous paragraph). 
Note this normalization affects the result for $C_{\parallel}$ 
but not the significant result for $A_{\parallel}$.   
Expressing the double-integral \Eq{eLqWW} with $k = (P{+}P_0)/2$ and $p=P{-}P_0$, 
and using the notation ${z=(c/T)}$, it reads  
\beq \label{eWkernel}
\tilde{\mathcal{W}}(r,r_0) \ \ = \ \ 
\nu \int_{-w/2}^{w/2} dp \, C e^{(i p/2) \left( r+r_{0} \right)} 
\int_{-\pi}^{\pi} \dfrac{dk}{2 \pi} \, e^{i k (r - r_0)} 
\exp \left[- z \sin{(p/2) \sin(k)} \right] 
\eeq
The inner integral may be written as ${\mathrm{I}_{r-r_0} [-z \sin(p/2)] ]}$,
however to calculate $A_{\parallel}$  in the high-temperature limit
it is enough to Taylor expand $z$ to second order.
The inner integral provides ``selection rules''. 
The zero order result gives a constant along the main diagonal of $\tilde{\mathcal{W}}$,
while the first order contributes to the near-neighbor hopping ($|r {-} r_0|=1$). 
The second order contributes both to next-near-neighbor hopping and to the main diagonal.

Including in $ \mathcal{L}^{(q)}$ also the $c \mathcal{L}^{(c)}$ term, 
and using the method described in \Sec{sec:perturbation}, one finds 
\begin{align}\label{eq:D-ell-ohmic}
A_{\parallel}(w) = 
\frac{w^2+16 \sin ^2\left(\frac{w}{2}\right)-\sin ^2(w)+4 w \sin (w)-4 \sin \left(\frac{w}{2}\right) (3 w+\sin (w))}{16 \left(w-2 \sin \left(\frac{w}{2}\right)\right) (w-\sin (w))}
-\frac{3 w-8 \sin \left(\frac{w}{2}\right)+\sin (w)}{32 \left(w-2 \sin \left(\frac{w}{2}\right)\right)}
\end{align}
The first term is the Ohmic result, while the second term is added to get the Boltzmann-corrected result.
The results for the X-coupling and for the S-coupling in \Eq{eA} are obtained for ${w=0}$ and ${w=2 \pi}$ respectively.

{\bf Boltzmann case for S/B.-- }
For the Boltzmann-corrected versions of the $S$ and $B$ one obtains 
\beq \label{eLqSB}
\tilde{\mathcal{W}}^{(S)}(r,r_0) \  &=& \  \nu_{S} \mathrm{I}_r(z/2) \mathrm{I}_{r_0}(-z/2) 
\\
\tilde{\mathcal{W}}^{(B)}(r,r_0) \  &=& \   2 \nu_{B} \mathrm{I}_r(z/2) \mathrm{I}_{r_0}(-z/2) 
+ \nu \left[ \mathrm{I}_{r+1}(z/2) \mathrm{I}_{r_0-1}(-z/2) + \mathrm{I}_{r-1}(z/2) \mathrm{I}_{r_0+1}(-z/2) \right]
\eeq
Expanding in $z=(c/T)$ the first order result for $\mathcal{L}^{(q)}$ is a ${q=0}$ version of the S/B dissipators 
that were presented in the main text \Eq{eLterms}. 
In the Boltzmann-corrected approximation both schemes acquire 
second-order terms ${-(3/16)(c/T)^2 \nu \ket{\pm 1}\bra{\pm 1}}$
that are required for the calculation of $D$.
For the $S$ coupling scheme one finds additional second-order terms that are needed for the calculations,
namely, ${-(3/32)(c/T)^2 \nu \ket{1}\bra{-1}}$ and  ${-(3/32)(c/T)^2 \nu \ket{-1}\bra{1}}$.

\clearpage



%

\end{document}